\documentclass[%
 aip,
 rsi,%
 amsmath,amssymb,
 reprint,%
]{revtex4-1}
\usepackage{natbib}
\usepackage{color}

\usepackage{graphicx}
\usepackage{dcolumn}
\usepackage{bm}

\usepackage{siunitx}

\begin{document}


\title{An optically-heated atomic source for compact ion trap vacuum systems}

\author{S. Gao}
\author{W. Hughes}
\affiliation{Department of Physics, University of Oxford, Clarendon Laboratory, Parks Rd, Oxford, OX1 3PU, UK}
\author{D. M. Lucas}
\affiliation{Department of Physics, University of Oxford, Clarendon Laboratory, Parks Rd, Oxford, OX1 3PU, UK}
\author{T. G. Ballance}
\affiliation{ColdQuanta UK, Oxford Centre for Innovation, Oxford, OX1 1BY, UK}
\author{J. F. Goodwin}
\email{joseph.goodwin@physics.ox.ac.uk}
\affiliation{Department of Physics, University of Oxford, Clarendon Laboratory, Parks Rd, Oxford, OX1 3PU, UK}

\date{\today}

\begin{abstract}
We present a design for an atomic oven suitable for loading ion traps, which is operated via optical heating with a continuous-wave multimode diode laser. The absence of the low-resistance electrical connections necessary for Joule heating allows the oven to be extremely well thermally isolated from the rest of the vacuum system, and for an oven filled with calcium we achieve a number density suitable for rapid ion loading in the target region with \SI{\sim200}{\milli\watt} of laser power, limited by radiative losses. With simple feedforward to the laser power, the turn-on time for the oven is less than \SI{20}{\second}, while the oven contains enough calcium to operate continously for many thousands of years without replenishment.
\end{abstract}

\maketitle

\section{Introduction}

Cold trapped atomic ions are an ideal platform for quantum computing beyond the fault tolerant threshold, having demonstrated the highest fidelity operations (preparation, readout, and single- and two-qubit gates) and the highest ratio of coherence time to gate period\cite{Ballance2016, Harty2014, Schafer2018, Sepiol2019} of any qubit candidate.
Although trapped ions qubits can be manipulated with exquisite precision, the translation of these techniques to a larger-scale processor capable of operating at the $\sim1000$-qubit level has yet to be demonstrated. The technical challenge of building and controlling a large system is significant and a number of approaches have been proposed to address this\cite{Kielpinski2002, Lekitsch2017, Nickerson2014}.

One approach to achieve the necessary scaling is to produce a large number of relatively simple quantum processor nodes, each of a size which permits high-fidelity operation, and to connect the nodes together to form a large quantum network. Entanglement between modules can be generated and distributed remotely via photonic interconnects\cite{Moehring2007, Stephenson2019}. By taking a modular approach the complexity of each node is reduced to a managable level, thereby transforming the scaling challenge from a problem of quantum physics to one of classical systems engineering and integration.

Construction of a 1000-qubit machine with this approach, assuming 10 application qubits per node, would require a network of 100 ion trap systems. Present-day proof-of-principal ion trap systems do not have particular size requirements, however in the construction of these larger networks the per-node footprint will become increasingly critical. The ultra-high vacuum system which houses the ion trap typically dictates the overall size of each node, and therefore pursuing compact and scalable technologies for this particular component is important.

A number of technical aspects of system design are improved when the vacuum system is scaled down, including reduced imaging system working distance, increased mechanical and thermal stability, and lower power requirements for magnetic field coils. On the other hand, there are aspects which become more challenging, primarily achieving a suitable quality of vacuum with limited pump rates, putting a strong imperative on making vacuum-side components as simple and clean as possible and minimizing localised heating of the system to reduce outgassing. One such in-vacuum component is the atomic source used to load the ion trap, which typically consists of a resistively heated oven connected to a pair of high-current vacuum feedthroughs. This approach to construction leads to a relatively complex and poorly thermally isolated assembly, increasing both the in-vacuum surface area and unwanted heating of the surrounding system. In this paper we demonstrate how optical heating of the oven mitigates these issues and demonstrate an effective miniaturised atomic source compatible with ultra-compact ion trap systems.

\section{Atomic sources in ultra-compact vacuum systems}

Loading of ion traps requires ionising atoms of the desired species within the trapping region, commonly through a two-photon ionization process\cite{DML2004}. A variety of methods are used experimentally to produce sufficient atomic density in the loading region, including collimated thermal beams\cite{Ballance2018}, laser ablation \cite{Hendricks2007}, and cold beams from magneto-optical traps (MOTs)\cite{Sage2012}.

Collimated thermal beams are the most widely used due to the simplicity with which they can be produced: a small tube filled with the target species can be heated with an electric current with high reliability and loaded with sufficient source metal to last for thousands of years of loading if used conservatively. A highly compact Joule-heated thermal source has been demonstrated\cite{Schwindt2016}, however high-current electrical feedthroughs remain a potential point of failure for ultra-compact systems. Furthermore, Joule heating is inherently inefficient as the low-resitivity electrical connections needed to deliver the heating current also act to conductively sink the generated heat, greatly increasing overall power requirements and making the integration of such sources into cryogenic systems unfeasible.

Laser ablation loading has gained popularity in recent years due to the low energy delivered per ion, the absence of high-current feedthroughs, and the lack of warm-up time allowing near-instantaneous ion loading\cite{Vrijsen2019}. However, this comes at the cost and complexity of including a high-intensity pulsed laser source and, in the case of many-node networked quantum computers, directing the emitted pulses to each node. The ablation plume produced is also typically less `clean' with direct ionisation processes occuring in parallel with the emission of neutral atoms, and care must be taken to avoid introducing contaminant ions to the trap.

MOT-based loading provides a clean, cold and well-controlled source of atoms on demand, but the considerable experimental overheads make it all but unfeasible for scaling to large networks of small traps.

In this paper we investigate an alternative approach to producing the atomic source, using an optically heated thermal source which combines the clean output associated with Joule-heated sources with greatly improved efficiency and a reduced complexity of construction.\color{black}

\section{Construction}

The oven consists of a stainless steel tube, of length \SI{10}{\milli\metre}, diameter \SI{2}{\milli\metre} and wall thickness \SI{0.1}{\milli\metre}. The oven is filled with approximately \SI{20}{\milli\gram} of calcium in granular form and both ends closed with stainless steel plugs, one of which has a \SI{0.75}{\milli\metre} aperture drilled through its centre. The oven is mounted on a framework of glass rods and folded steel supports, which ensure thermal conductivity between the oven and the surrounding room temperature vacuum system is extremely low.

The completed assembly is mounted in a custom built ultra-compact vacuum system using ColdQuanta's 'Channel Cell' technology, of volume \SI{<10}{\centi\metre^{3}}, with an integrated ion pump. The oven is directed through a second \SI{0.75}{\milli\metre} aperture fixed to an internal silicon partition of the chamber, reducing deposition of calcium outside the target region. This target region consists of a fiducial in the form of a silicon wafer with a $5\times\SI{0.5}{\milli\metre}$ slot cut through it, representing the intended location of the ion trap.

Heating of the atomic oven is achieved via a multimode continuous-wave diode laser near \SI{780}{\nano\metre}, capable of up to \SI{7}{\watt} output power, but typically operated at \SIrange{100}{500}{\milli\watt}. The heating laser is focused and directed onto the oven tube with ex-vacuo bulk optics, targeting a small region of the tube which is intentionally darkened to increase absorption. For improved performance, the remainder of the oven tube can be electroplated in gold to reduce the radiative emissivity, although this has not been performed for the results reported here.

After filling the oven under atmospheric conditions, the calcium granules are covered in a thick oxide layer, which prevents rapid evolvement of calcium vapour when heated to the relatively low temperatures ($\SI{<500}{\kelvin}$) required for ion trap loading, making the response slow and unreliable. To overcome this, after evacuating the system the oven must be `cracked' by briefly operating at a much higher temperature ($\SI{\sim700}{\kelvin}$), which fractures the oxide layer, exposing fresh and readily evaporable calcium metal. Unfortunately, this process is accompanied by a very high-flux atomic beam, which can rapidly coat ion trap electrodes with calcium metal, leading to significant and irreparable damage. To prevent this occuring, the second aperture is preceded by a mechanical shutter which can be closed until after the `cracking' process is complete, and opened in-vacuo via magnetic actuation.

\begin{figure}
\includegraphics[width=\columnwidth]{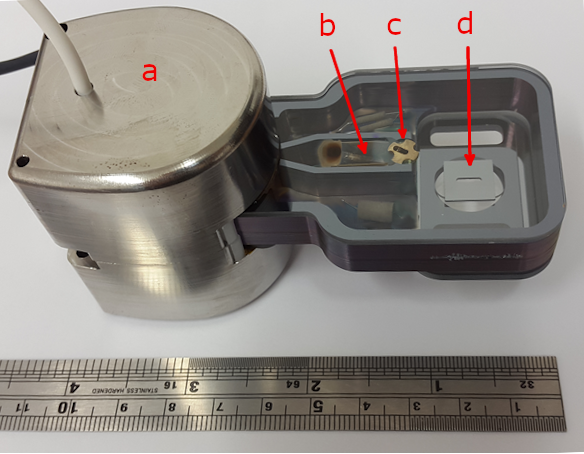}
\caption{`MITAS' ultra-high vacuum system with (a) integrated Penning cell ion pump; (b) optically heated calcium oven; (c) magnetically actuated oven shutter; (d) target region with dummy trap. Internal volume of cell is \SI{<10}{\centi\metre^3}.}
\label{fig:cell}
\end{figure}

\section{Characterisation of Oven Performance}
To validate the performance of the oven, we evaluate the dynamic and equilibrium number density of atoms produced within the target region versus heating laser power, via spectroscopic measurements. We target a nominal number density of $\SI{100}{\centi\metre^{-3}}$, sufficient to achieve ion loading within a few seconds in typical trap geometries with $\sim\SI{1}{\milli\watt}$ of power applied via the second-stage ionisation laser~\cite{DML2004}. We also infer the temperature of the oven from the measured number density and system geometry, and use this to produce a simple thermal model of the oven and give a prediction of its operating lifetime under normal conditions.

\subsection{Measurement of atomic flux}
The oven tube is optically heated by the heating laser, producing a wide beam of neutral calcium atoms from the first of the two \SI{0.75}{\milli\metre} collimation apertures, the second of which selects a narrow beam of atoms directed towards the target region. We detect the presence and number density of neutral calcium atoms via resonance fluorescence spectroscopy on the $4\textrm{s}^2{^1\textrm{S}_0}\to4\textrm{s}4\textrm{p}^1\textrm{P}_1$ transition. The atomic beam passing through the target region interacts with the beam from a \SI{423}{\nano\metre} single-mode external-cavity diode laser, which is focused to a waist of $1/e$ radius $w_0\approx\SI{40}{\micro\metre}$ and aligned with wavevector orthogonal to the neutral atom flux, minimising Doppler effects. The resonance fluoresence emitted by the atoms passing through this target region is collected by an $\textrm{NA}\sim0.3$ lens and imaged with magnification $m\sim10$ to a photon-counting photomultiplier tube (PMT) for detection. A \SI{500}{\micro\metre} aperture is placed immediately before the PMT, such that detection is limited to photons emitted from a $(\lesssim\SI{50}{\micro\metre})^3$ target region where the spectroscopy laser and atomic beam intersect.

\begin{figure}
\includegraphics[width=\columnwidth]{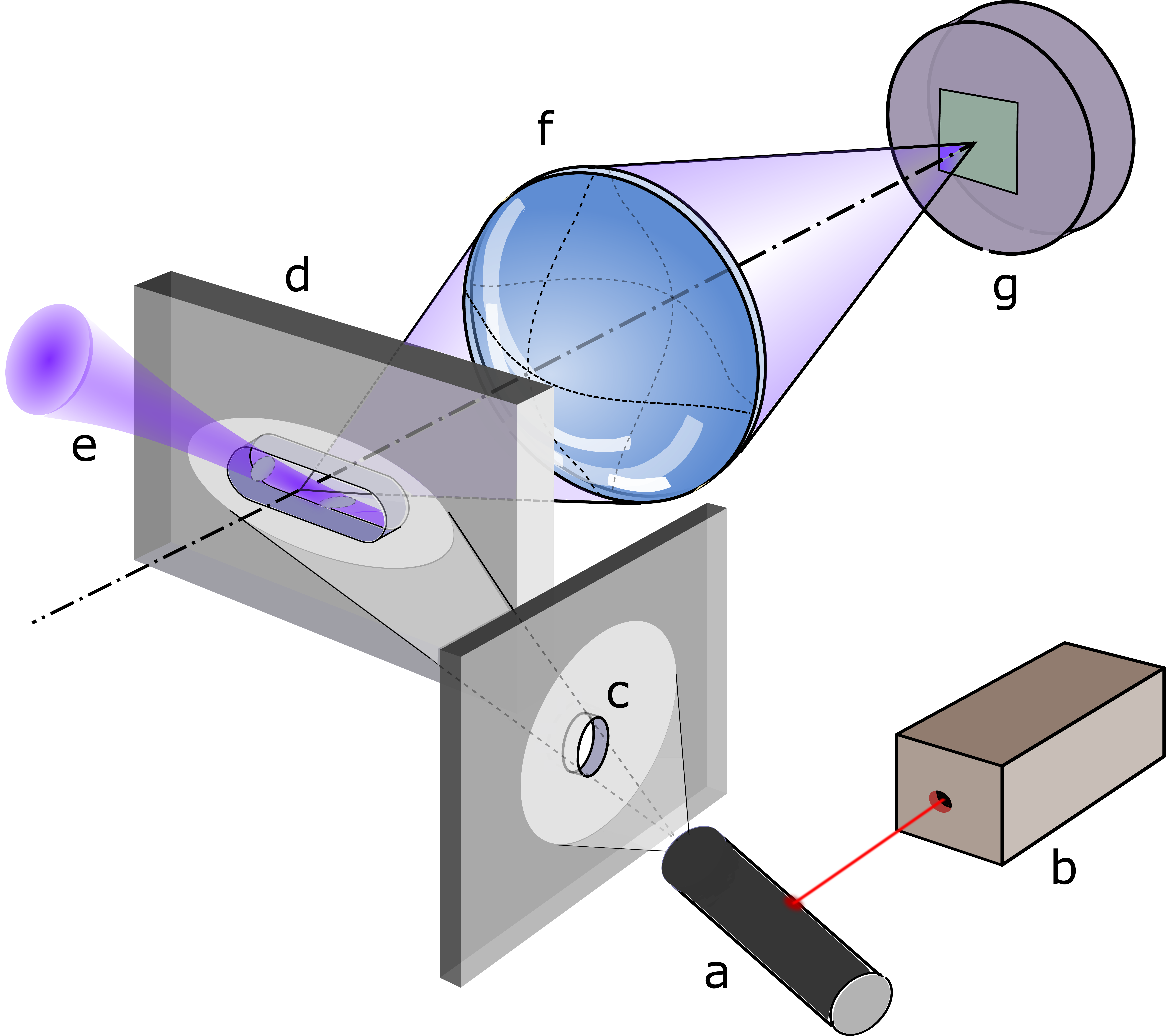}
\caption{Schematic diagram of apparatus for optical heating of calcium oven and resonance fluorescence spectroscopy of emitted atomic beam. (a) Calcium-filled stainless steel oven tube; (b) \SI{780}{\nano\metre} optical heating laser; (c) beam of calcium atoms passing through second collimation aperture; (d) silicon wafer target, approximating a microfabricated wafer-type ion trap chip; (e) \SI{423}{\nano\metre} spectroscopy laser beam aligned with wavevector orthogonal to the neutral atom flux; (f) $\textrm{NA}\sim0.3$ collection lens, imaging with magnification $m\sim10$ to (g) photon-counting PMT detector.}
\label{fig:apparatus}
\end{figure}

From the imaging system photon detection efficiency, the interaction and fluoresence collection volumes and the physics of the laser-atom interactions, we can convert the measured photon count rate to a number density of atoms in the target region.

Throughout these experiments, the intensity of the excitation laser is kept much smaller than the saturation intensity\footnote{Note that for experimental simplicity, we do not apply an external magnetic bias field so the quantisation axis is not well defined. Similarly, the polarisation of the spectroscopy laser, while stable throughout the experiments, is not calibrated. For this reason we assume isotropic emission and a saturation intensity defined for unpolarised light. These assumptions imply a significant uncertainty in our measurements of absolute count rate, but do not affect our ability to ascertain the response time nor (due to the very strong dependence of vapour pressure upon oven temperature), the thermal efficiency of the atomic source, which are the only aims of this experiment.},
\begin{equation}
    I_{sat} = \frac{\hbar\omega_0^3 \Gamma}{4 \pi c^2},
\end{equation}
where $\omega_0=2\pi\times\SI{707}{\tera\hertz}$ is the frequency of the spectroscopy transition and $\Gamma=2\pi\times\SI{35}{\mega\hertz}$ is the transition linewidth.
The total scattering rate, for an ensemble of atoms with Doppler-shifted transition frequencies at detuning $\Delta$ from the excitation field frequency, is then given by
\begin{equation}
    R = \int\int_V I(r,z) \kappa(\Delta) n(\Delta)\textrm{d}V\textrm{d}\Delta
\label{scattering_rate}
\end{equation}
where
\begin{equation}
\kappa(\Delta) \approx \frac{\Gamma}{2I_{sat}(1+4\frac{\Delta^2}{\Gamma^2})}
\end{equation}
and $I(r,z)$ is the excitation field intensity as a function of radial and axial position within the spectroscopy beam, $n(\Delta)\textrm{d}\Delta$ is the number density of calcium atoms with Doppler-induced transition frequency shifts $\Delta+\textrm{d}\Delta$ from the laser frequency, and the inner integral is taken over the volume $V$ of the region imaged onto the detector.

The excitation field is orthogonal to the atomic beam axis, so the only Doppler shifts are those due to the divergence of the atomic beam across the interaction region, with
\begin{equation}
\Delta(\theta,T) \approx 2\pi\theta \left(\frac{\bar{v}(T)}{\lambda}\right),
\end{equation}

where $\theta$ is the small angle between the atom velocity and the axis of the atomic beam in the plane containing the axis of the excitation beam, $\bar{v}(T) = \sqrt{8k_B T/m\pi}$ is the mean speed of atoms with mass $m$ in a beam emitted from a gas at temperature $T$, and $\lambda$ is the wavelength of the excitation field.

For the purposes of estimating the Doppler shift number density $n(\Delta)$, we consider the target region to be point-like, a good approximation given its size and distance from the source. The direction of the velocity of an atom reaching the target region is thus associated with the region of the source aperture from which it originates. As the collimation aperture is circular, we calculate the Doppler-shift spectral number density
\begin{equation}
    n(\Delta) = N*p(\Delta)=N*\frac{2}{\pi \Delta_{max}}\sqrt{1-\left(\frac{\Delta}{\Delta_{max}}\right)^2},
\end{equation}
where $N$ is the total number density in the target region and the $p(\Delta)$ is the Doppler shift probability distribution. $-\Delta_{max}(T)<\Delta<\Delta_{max}(T)$, where $\Delta_{max}(T)$ is determined by the temperature and the collimation of the atomic beam. 

In our set-up, $\theta_{max} = 0.044$, given by the first aperture diameter (\SI{0.75}{\milli\metre}) and its distance to the target region (\SI{17}{\milli\metre}).

We can evaluate the integrals separately for each term in Eqn.~\ref{scattering_rate}. The intensity is near-constant with $z$ over the length of the interaction region, so evaluating the volume integral (see~App.~\ref{app:A} for details) we find
\begin{equation}
    \int_V I(r,z)\textrm{d}V = 2\pi\int_{-L/2}^{L/2}\int_0^\infty I(r,z)\textrm{d}r\textrm{d}z= L P_{423},
\label{eq: intensity volume integral}
\end{equation}
where $L$ is an effective length of the imaging system collection volume in the direction of the laser beam and $P_{423}$ is the total optical power. We therefore find
\begin{equation}
\begin{split}
    R &=L P_{423} \int\kappa(\Delta)n(\Delta)\textrm{d}\Delta \\
       &= L P_{423} N \tilde{\kappa}(T)
\end{split}
\end{equation}
where $N(T)$ is the number density and  
\begin{equation}
\begin{split}
\tilde{\kappa}(T) =& \int_{-\Delta_{max}(T)}^{\Delta_{max}(T)}\kappa(\Delta)p(\Delta)\textrm{d}\Delta\\
=&\frac{\Gamma^3}{4I_{sat}\left[\Delta_{max}(T)\right]^2}\left(\sqrt{1+\frac{4\left[\Delta_{max}(T)\right]^2}{\Gamma^2}}-1\right).
\end{split}
\end{equation}

The photon count rate is $C = \eta_{d}\eta_{g}R$, where $\eta_d\approx 25\%$ is the detection efficiency including losses in the optics and $\eta_g \approx 1.7\%$ is the geometric collection efficiency calculated from the numerical aperture ($\textrm{NA}\approx0.3$) of the photon-collecting lens.
The number density $N(T)$ can therefore be determined from the measured count rate and the temperature of the beam:
\begin{equation}
    N(T) = \frac{C}{\tilde{\kappa}(T) P_{423} L\eta_d\eta_g}.
    \label{count to number density}
\end{equation}

\subsection{Thermal performance}

\begin{figure}
\includegraphics[width=\columnwidth]{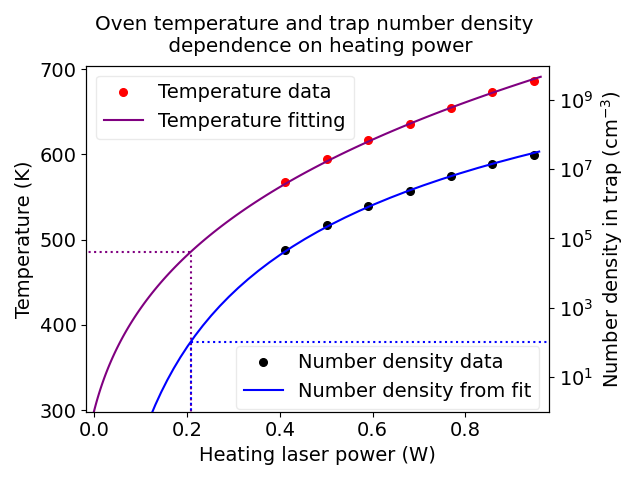}
\caption{Dependence of oven output on optical heating power. We measure steady-state number density in the target region as a function of incident power, and convert this to a projected oven temperature via the geometric properties of the system and standard calcium vapour pressure curves. We fit a simple thermal model of the oven to this data, finding that radiative processes dominate the power loss. Using the same known properties of the system to convert the temperature fit to number density allows us to find the power necessary to achieve the target number density (\SI{100}{\cm^{-3}}) in the trap region, which is too low to easily measure via resonance fluorescence spectroscopy. We predict that \SI{200}{\milli\watt} of optical power is sufficient to run the oven at the desired level of flux.}
\label{fig:temp_den}
\end{figure}

The temperature of the oven could be determined by scanning the laser to measure the Doppler broadening, but the intentionally near-Doppler-free geometry of the system means this provides limited sensitivity. However, the very strong dependence of vapour pressure on temperature means that by comparing the measured number density with that predicted to be found within the interaction region, we are able to infer the oven temperature to within a few kelvin, even when considering the substantial uncertainty in $N$ due to the uncalibrated polarisation.

The atomic vapour pressure within the oven is given by
\begin{equation}
    p(T) = N_o(T)k_B T,
\end{equation}

where $N_o$ is the number density inside the oven and $T$ the oven temperature. The vapour pressure can be obtained from standard tables. The atomic flux density at the hole is $\Phi = N_{o}(T)\bar{v}(T)/4$, where $N_{o}(T)$ is the number density in the oven, $\bar{v} = \sqrt{8k_B T/m\pi}$ is the mean speed of atoms in the oven (equal to that in the emitted beam). The atomic flux density on the sphere $r$ away from the source is $\phi(r,\beta) = \phi_0(r) \text{cos}\beta$, where $\beta$ is the angle to the atomic beam axis and flux conservation of the half sphere with radius $r$ determines axial flux density $\phi_0(r)$:
\begin{equation}
    \Phi A_o= \int_0^{\pi/2} \phi(r,\beta) (2 \pi r\text{sin}\beta) r d\beta = \phi_0(r) \pi r^2
    \label{oven_flux}
\end{equation}
where $A_o$ is the area of the oven aperture. The flux density along the axis is also related to the target region number density $N$ by $\phi_0(r_o)=\bar{v}N$, where $r_o$ is the distance between the oven aperture and the target region. The target region number density is then:
\begin{equation}
        N(T) = \frac{\phi_0(r_o)}{\bar{v}(T)}=\frac{N_o(T)A_o}{4\pi r_o^2}  =\frac{p(T)A_o}{4 \pi k_B T r_o^2 }
    \label{intrinsic number density}
\end{equation}

Using equations (\ref{intrinsic number density}) and (\ref{count to number density}), we can calculate the number density in the target region and the temperature of the oven from the measured photon count rate. In Fig.~\ref{fig:temp_den} we show how steady-state number density and oven temperature vary as a function of optical power.
Considering in particular how the temperature varies with power, we can fit a thermal model of the oven and extract estimates of its radiative emissivity $\epsilon_{rad}$ and the thermal conductivity $C$ of its mechanical supports. For an oven at temperature $T$ mounted in a vacuum system at room temperature $T_0 = \SI{300}{\kelvin}$, the total heat loss power is:
\begin{equation}
    P_{loss} = C (T-T_0) + \epsilon_{rad}\sigma S (T^4-T_0^4)
\end{equation}
where $\sigma$ is the Stefan-Boltzmann constant and $S$ is the surface area of the oven.
The fraction of heating laser power absorbed $\epsilon_{abs}$ may differ from the total surface emissivity (due to both the difference in wavelength and local variation in surface roughness or composition). From energy conservation we therefore find:
\begin{equation}
    P_{laser} = \frac{P_{loss}}{\epsilon_{abs}}=\frac{C}{\epsilon_{abs}} (T-T_0) + \frac{\epsilon_{rad}}{\epsilon_{abs}}\sigma S (T^4-T_0^4).
\end{equation}
Fitting this to the data in Fig.~\ref{fig:temp_den} we find the contribution of conductive losses is insignificant and we are left with a purely radiative model
\begin{equation}
    T\approx\left[\left(\frac{\epsilon_{abs}}{\epsilon_{rad}}\right)\frac{P_{laser}}{S\sigma}\right]^{1/4}
\end{equation}
with a single fit parameter representing radiative heating efficiency, which we find to be
\begin{equation}
    \frac{\epsilon_{rad}}{\epsilon_{abs}} = 1.22.
\end{equation}
From our fitted thermal model and the results in Eqns.~\ref{intrinsic number density} and \ref{count to number density}, we can then predict how number density will vary with laser power, as shown by the blue line in Fig.~\ref{fig:temp_den}. This allows us to extrapolate from our measured data to number densities which would not produce enough fluorescence to permit direct detection; achieving our target number density of \SI{100}{\centi\metre^{-3}} -- corresponding to an average number of atoms in the loading region at any one time of $<0.0001$ -- is predicted to require an oven temperature of \SI{\sim485}{\kelvin} and a heating power of \SI{200}{\milli\watt}.

\subsection{Temporal response}

\begin{figure}
\includegraphics[width=\columnwidth]{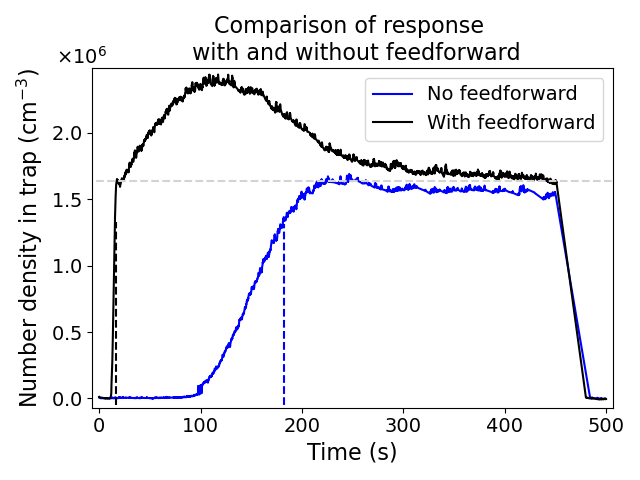}
\caption{Temporal response of number density in loading region after heating laser turn-on, with and without feedforward to optical power. Operating at constant heating power, the oven takes \SI{\sim3}{\minute} to reach 80\% of its steady state output; with a simple two-level turn-on profile this response time can be reduced to \SI{\sim17}{\second}.}
\label{fig:temporal}
\end{figure}

We now consider the temporal response of the oven immediately after the heating laser is switched on. The blue data series shown in Fig.~\ref{fig:temporal} shows the number density measured in the trap region after applying \SI{500}{\milli\watt} of laser power from $t=0$ to $t=$ \SI{440}{\second}. The number density begins to rise after about a minute of heating, and reaches 80\% of its steady-state value ($\SI{\sim16e6}{\centi\metre^{-3}}$) after \SI{182}{\second}. To improve the response time of the source, we apply very simple feedforward in the form of two-level heating profile, with \SI{3}{\watt} of laser power applied from $t=0$ to \SI{5}{\second}, before being reduced to \SI{500}{\milli\watt} from $t=$ \SI{5}{\second} to \SI{440}{\second}. This results in the number density response shown in black in Fig.~\ref{fig:temporal}, with 80\% of the steady-state value reached in \SI{17}{\second}. More sophisticated power profiles could doubtless achieve even faster response and less overshoot, but in practice it is likely that this simple approach is sufficient for most applications.

\subsection{Predicted lifetime}

Using, Eq.~\ref{intrinsic number density}, the total flux of atoms leaving the oven, given in Eq.~\ref{oven_flux} is
\begin{equation}
J_{atom}= \Phi A_o=N(T)\bar{v}(T)\pi r_0^2.
\end{equation}
Operating at \SI{485}{\kelvin}, sufficient to produce the $\SI{100}{\centi\metre^{-3}}$ target number density, the oven emits atoms at a rate of $J_{atom}=46\times10^6\SI{}{\second^{-1}}$. The \SI{20}{\milli\gram} of calcium in the oven consists of $\sim3\times10^{20}$ atoms, giving an expected lifetime under continuous operation of 200,000 years.

\section{Conclusion and outlook}

We have produced a compact optically heated atomic oven. By avoiding the increased thermal conductive losses associated with the wiring required for electrical heating, the oven may be extremely well thermally isolated from the surrounding vacuum system, and we find \SI{200}{\milli\watt} of optical power heats the oven to \SI{485}{\kelvin}, producing atomic flux density suitable for rapid loading of a ion trap. The power required could be further reduced through gold electroplating of the oven tube. Running at high steady-state flux levels and with a maximum of \SI{3}{\watt} of laser power, we achieve 80\% of equilibrium flux within \SI{17}{\second} with simple two-level feedforward to the laser power; an even faster response is likely possible when operating at the much lower flux levels required for ion loading, by using a higher optical power, or with a smaller oven with lower thermal mass.

The extremely high lifetime of the current source implies the mass of calcium in the oven, and thus its size, could be greatly reduced without any meaningful impact. Looking ahead, we are investigating far smaller optically heated sources, integrated within the structure of the trap itself. By also moving the source much closer to the target region, the overall flux required to reach a suitable number density for loading is greatly reduced, and the source mass may be reduced further still. Combined with measures to reduce radiative losses, the necessary heating power could likely be reduced by an order of magnitude or more, reducing the heating laser requirements and presenting the possibility of operation within a cryogenic system.

While such reductions in mass also provide the potential for reduced turn-on time, another approach worthy of consideration is to forgo the question of temporal response altogether by running the oven continuously at a very low flux. For suitable choice of oven materials and location within a multi-zone trap, this may be feasible without significant reduction in vacuum quality in the qubit operation regions. Loading would then be achieved by applying a short pulse of both photoionisation lasers, with ion production rate limited only by the second stage power available. This approach would also bring the benefit of reducing temporal fluctuations of the temperature throughout the trap, which can introduce problematic drifts in the electric properties and micromotion compensation.

In summary, we believe that efficient, optically heated thermal atomic sources provide a promising solution to the problem of ion loading in compact ion trap vacuum systems, where the low power dissipation and simplicity of construction and assembly represent considerable benefits.

\section*{Acknowledgments}
The authors thank Jinen Guo for assistance with the calibration of the imaging system. This work was supported by Innovate U.K. Project \#104159 (``MITAS'') and the U.K. EPSRC “Networked Quantum Information Technology” Hub (Grant No. EP/M013243/1).

\vspace{1cm}

\appendix
\section{Calculation of effective fluorescence collection region volume} \label{app:A}
In equation \ref{eq: intensity volume integral}, the total integrated intensity of the collected volume $\int_V I(r,z)\textrm{d}V$ is converted to the product of the power $P_{423}$ and an effective length $L$. This effective length $L$ is determined by the area over which the excitation beam and the collection overlap in space (the atomic beam is sufficiently wide that its properties can be assumed constant over the relevant region).

The beam has a $1/e$ waist of $w_0=\SI{40}{\micro\metre}$, and thus a Rayleigh range of $z_0=\SI{12}{\milli\metre}$. This is much longer than the extent of the beam in the collection region, and thus the beam radius will be considered constant over the entire collection region.

The imaging system has a magnification of $\approx$10 through a \SI{500}{\micro\metre} diameter aperture before the PMT. Therefore counts can be collected from up to \SI{25}{\micro\metre} away from the imaging axis in the imaging plane. The imaging system is assumed to have infinite depth of field because the depth of field is much larger than the imaging axis length over which atoms are excited by the laser

\begin{figure}
\includegraphics[width=\columnwidth]{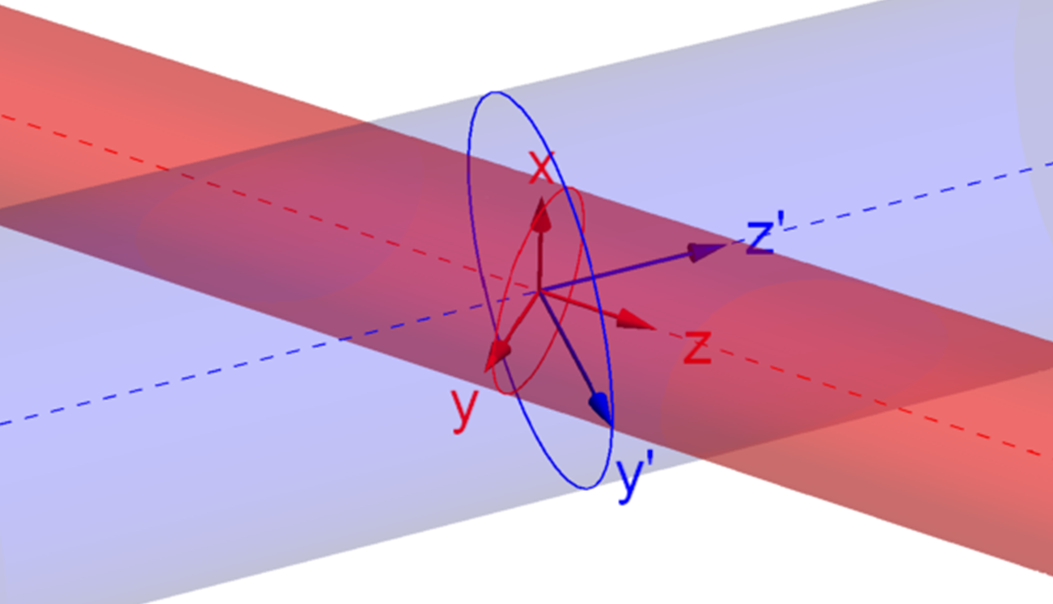}
\caption{Diagram of the coordinate system geometry used to calculate the effective length $L$ of the imaging system. The collection region is shown in red and the excitation beam (shown here as a cylindrical contour) is blue. The target region is formed at the overlap of the excitation beam and the collection.}
\label{fig: appendix collection volume}
\end{figure}

With these assumptions, the volume integral of intensity in the collection region $\int_V I(r,z)\textrm{d}V$ can be calculated by setting up two coordinates systems according to figure \ref{fig: appendix collection volume}. The coordinates $x$, $y$ and $z$ are associated with the imaging system, with $z$ defining the imaging axis and the origin in the focal plane. The coordinates $x'$, $y'$ and $z'$ are associated with the excitation beam, with $z'$ along the beam axis and $x'$ and $y'$ in the transverse plane. The two coordinate systems share a common origin. Without loss of generality, $x'$ is chosen to coincide with $x$.

The intensity of the excitation field in its coordinate system can be written
\begin{equation}
I(x',y',z') = \frac{2P_{423}}{\pi w_0^2} e^{-\frac{2\left(x'^2+y'^2\right)}{w_0^2}},
\end{equation} 
where there is no $z'$ dependence because the Rayleigh range is assumed large. To transform into the collection coordinates, we use $x'=x$ and $y'=y\cos\alpha+z\sin\alpha$ with $\alpha = 45^{\circ}$ the angle between the excitation beam axis and imaging axis. Finally, we convert the transverse coordinates $x$, $y$ to polar coordinates $\rho$, $\phi$ using $x=\rho \cos\theta$ and $y=\rho \sin\theta$. This leaves the integral of equation \ref{eq: intensity volume integral} as
\begin{widetext}
\begin{equation}
    \int_V I(r,z)\textrm{d}V = \frac{2P_{423}}{\pi w_0^2}\int_{-\infty}^{\infty}\int_{0}^{\rho_m}\int_{0}^{2\pi} e^{-\frac{2}{w_0^2}\left(\left(\rho\cos\phi\right)^2+\left(\rho\cos\alpha\sin\phi + z\sin\alpha\right)^2\right)}\rho \, d\phi \, d\rho \, dz,
\end{equation}
\end{widetext}

where $\rho_m =\SI{25}{\micro\metre}$ is the radial extent of the collection region from the imaging axis. Evaluated numerically, this leads to $\int_V I(r,z)\textrm{d}V = L P_{423}$ with $L=\SI{46.4}{\micro\metre}$.

\end{document}